\documentclass[aps,preprint]{revtex4}
\usepackage{amssymb}
\usepackage{amsfonts}
\usepackage{amsmath}
\usepackage{graphicx}
\usepackage{dcolumn}
\usepackage{bm}

\input{tcilatex}

\begin{document}

\title{Muon spin rotation study of the intercalated graphite superconductor
CaC$_{6}$ at low temperatures}
\author{F. Mur\'{a}nyi$^{1}$, M. Bendele$^{1,2}$, R. Khasanov$^{2}$, 
Z. Guguchia$^{1}$, A. Shengelaya$^{3}$, C. Baines$^{2}$ and
H. Keller$^{1}$ }
\affiliation{$^{1}$Physik-Institut der Universit\"{a}t Z\"{u}rich, Winterthurerstrasse
190, CH-8057 Z\"{u}rich, Switzerland\\
$^{2}$Laboratory for Muon Spin Spectroscopy, Paul Scherrer Institute,
CH-5232 Villigen PSI, Switzerland\\
$^{3}$Department of Physics, Tbilisi State University, Chavchavadze 3,
GE-0128 Tbilisi, Georgia}
\date{\today }

\begin{abstract}
Muon spin rotation ($\mu $SR) experiments were performed on the intercalated
graphite CaC$_{6}$ in the normal and superconducting state down to 20 mK. In
addition, AC magnetization measurements were carried out resulting in an
anisotropic upper critical field $H_{\text{c2}}$, from which the coherence
lengths $\xi _{ab}\left( 0\right) =36.3(1.5)$ nm and $\xi _{c}\left(
0\right) =4.3(7)$ nm were estimated. The anisotropy parameter $\gamma _{H}=$ 
$H_{\text{c2}}^{ab}/H_{\text{c2}}^{c}$ increases monotonically with
decreasing temperature. A single isotropic $s$-wave description of
superconductivity cannot account for this behaviour. From magnetic field
dependent\ $\mu $SR experiments the absolute value of the in-plane magnetic
penetretion depth $\lambda _{ab}(0)=78(3)$ nm was determined. The
temperature dependence of the superfluid density $\rho _{s}\left( T\right) $
is slightly better described by a two-gap than a single-gap model.
\end{abstract}

\maketitle

\address{
}%

\section{\protect\bigskip Introduction}

The field of graphite intercalation compounds (GICs) gained attention after
the discovery of the superconductor CaC$_{6}$ with a rather high value of
the superconducting transition temperature $T_{\text{c}}\simeq $ 11.5 K \cite%
{weller_nature}. Superconductivity in GICs was first reported in the
potassium-graphite compound KC$_{8}$ with $T_{\text{c}}\simeq $ 0.14 K \cite%
{Hannay_KC8}. Until the discovery of the superconductor CaC$_{6}$ the
highest $T_{\text{c}}$ was observed in KTl$_{1.5}$C$_{5}$ with $T_{\text{c}%
}\simeq $ 2.7 K, synthesized under ambient pressure \cite{wachnik}. However,
high pressure synthesis was found to increase $T_{\text{c}}$ up to 5 K in
metastable compounds such as NaC$_{2}$ and KC$_{3}$ \cite{Belash,Avdeev}.
According to experimental and theoretical work CaC$_{6}$ can be described as
a classical BCS superconductor with a single isotropic gap\ $\Delta
_{0}\simeq 1.7$ meV \cite{STSpaper,LamuraPRL}. The upper critical field $H_{%
\text{c2}}$ shows a remarkable anisotropy with zero temperature values $\mu
_{0}H_{\text{c2}}^{c}(0)\simeq 0.3$ T and $\mu _{0}H_{\text{c2}%
}^{ab}(0)\simeq 1.9$ T, for the external field $H$\ parallel to the $c$-axis
or parallel to the $ab$-plane, respectively \cite{critical_field}. A recent
ARPES study indicated the existence of a possible second superconducting gap 
\cite{arpes} with a small zero temperature value $\Delta _{0,2}\simeq 0.2(2)$
meV. Tunneling experiments \cite{pointcontact} gave strong indications for
the existense of a superconducting anisotropic $s$-wave gap in CaC$_{6}$,
providing different zero temperature gap values for changing current
injection directions, $\Delta _{0}^{c}\simeq 1.7$ meV and $\Delta
_{0}^{ab}\simeq 1.44$ meV. The muon-spin rotation ($\mu $SR) technique is a
powerful method to characterise the superconducting gap symmetry in
superconductors \cite{Rustem1}. $\mu $SR measurements down to very low
temperatures may allow to access a possible second small gap, which should
manifest itself in the low temperature superfluid density in terms of an
inflection point \cite{Rustem1}. A recent $\mu $SR study \cite{DiCastro}
supports a single gap isotropic $s$-wave description of superconductivity in
CaC$_{6}$, although the low temperature region was not studied. In this work
we present extended measurements down to 20 mK and investigate possible
order parameter symmetries (single-gap isotropic $s$-wave, single-gap
anisotropic $s$-wave, two-gap isotropic $s$-wave) by means of $\mu $SR\
experiments.

\section{Experimental results}

CaC$_{6}$ samples were prepared from Highly Oriented Pyrolytic Graphite
(HOPG, Structure Probe Inc., SPI-2 Grade) and calcium metal (Sigma Aldrich,
99.99\% purity) by means of the molten alloy method. The detailed
preparation procedure is described elsewhere \cite{Emery_synthesis}. In
order to reduce the melting point of the calcium, lithium (Sigma Aldrich,
99.9\% purity) was added. The atomic ratio 3:1 of calcium and lithium was
measured in a stainless steel reactor under argon atmosphere. The molten
alloy was cleaned in the glove-box. After immersing the graphite pieces the
reactor was tightly sealed. Intercalation at 350 $^{0}$C was carried out in
an industrial furnace for 30 days. The resulting sample's color is gold and
the surface possesses a shiny metallic lustre. The final sample is a
superconductor with an onset temperature $T_{\text{c}}\simeq 11.2$ K, in
reasonable agreement with earlier studies \cite%
{weller_nature,STSpaper,LamuraPRL}.

\begin{figure}[th]
\includegraphics[width=0.9\hsize]{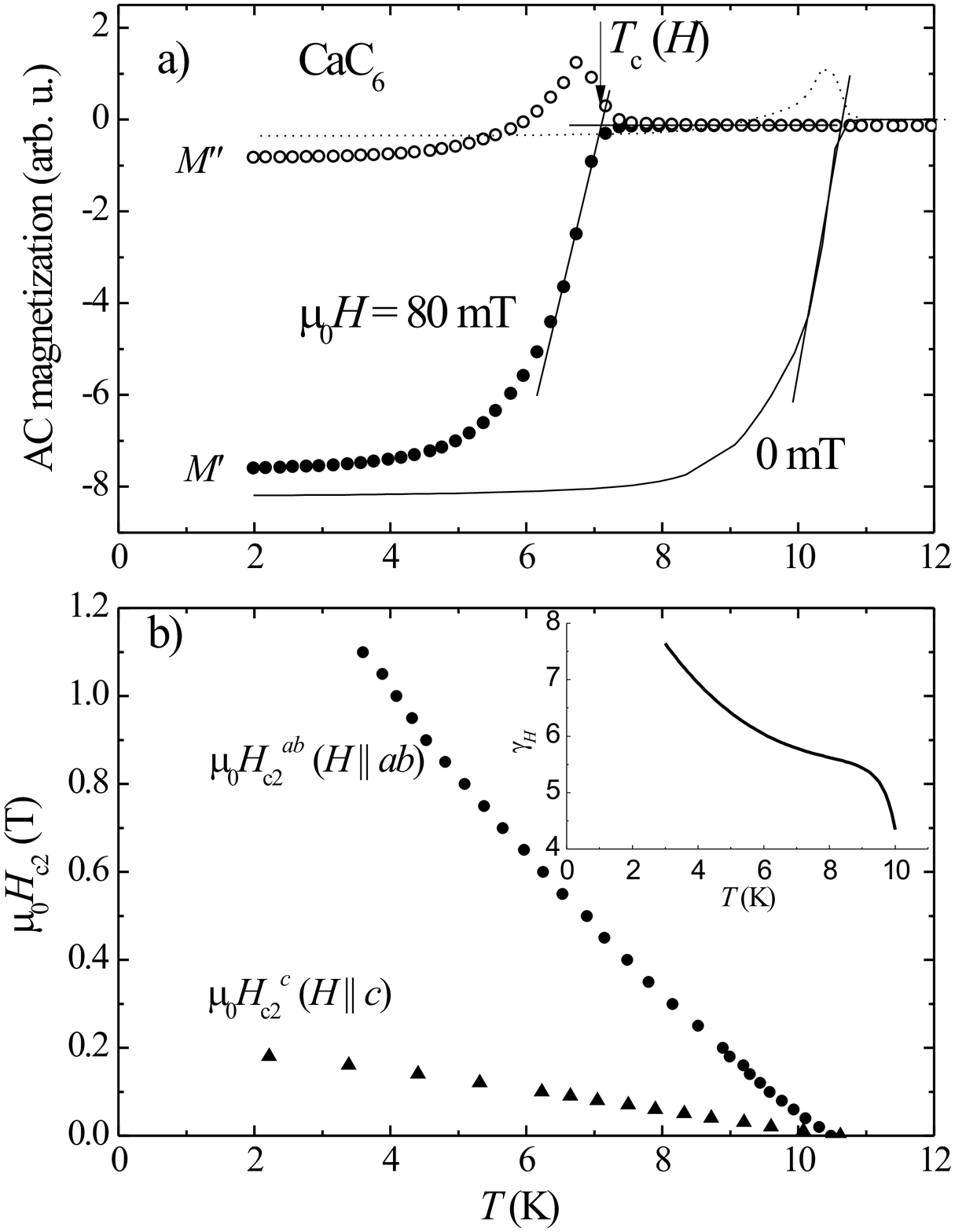}
\caption{a) AC magnetization curves of CaC$_{6}$ taken at $\protect\mu %
_{0}H= $ 80 mT and 0 mT. $M^{\prime }$ and $M^{\prime \prime }$ denote the
real and the imaginary part of the AC magnetization, respectively. The arrow
indicates the onset criterion to determine $T_{\text{c}}(H)$ (crossing of
the two lines). b) Temperature dependence of the upper critical fields $H_{%
\text{c2}}^{c}$ and $H_{\text{c2}}^{ab}$ for the external magnetic field $H$%
\TEXTsymbol{\vert}\TEXTsymbol{\vert}$c$ and $H$\TEXTsymbol{\vert}\TEXTsymbol{%
\vert}$ab$, respectively, as determined by AC magnetization. The inset shows
the calculated upper critical field anisotropy $\protect\gamma _{H}=H_{\text{%
c2}}^{ab}/H_{\text{c2}}^{c}$ as a function of temperature.}
\end{figure}
To characterise our sample we carried out detailed AC magnetization
measurements presented in Fig.~1. \textbf{The AC component of the applied
magnetic field was 1 mT at 1 kHz.} Figure 1a shows typical AC magnetization
curves at 80 mT (real $M^{\prime }(T)$ and imaginary $M^{\prime \prime }(T)$
part), the straight lines indicate the onset criterion to determine $T_{%
\text{c}}$. The temperature dependence of the upper critical field
components $H_{\text{c2}}^{c}$ ($H$\TEXTsymbol{\vert}\TEXTsymbol{\vert}$c$)
and $H_{\text{c2}}^{ab}$ ($H$\TEXTsymbol{\vert}\TEXTsymbol{\vert}$ab$) are
displayed in Fig.~1b. The values of the $H_{\text{c2}}^{c}$ and $H_{\text{c2}%
}^{ab}$ were determined from AC magnetization using the procedure
illustrated in Fig.~1a. The upper critical field is anisotropic and follows
the temperature dependence as reported in Refs. \cite{critical_field} and 
\cite{Cubitt_n}. In the framework of the anisotropic Ginzburg-Landau theory
the upper critical field for $H$\TEXTsymbol{\vert}\TEXTsymbol{\vert}$c$ and $%
H$\TEXTsymbol{\vert}\TEXTsymbol{\vert}$ab$ is given by

\begin{eqnarray}
\mu _{0}H_{\text{c2}}^{ab}\left( 0\right) &=&\frac{\Phi _{0}}{2\pi \xi
_{ab}\left( 0\right) \xi _{c}\left( 0\right) },  \notag \\
\mu _{0}H_{\text{c2}}^{c}\left( 0\right) &=&\frac{\Phi _{0}}{2\pi \xi _{ab}^{%
\text{2}}\left( 0\right) },
\end{eqnarray}%
where $\Phi _{0}=h/2e=2.07\cdot 10^{-15}$ Tm$^{2}$ is the flux quantum, and $%
\xi _{ab\text{,}c}$ are the corresponding coherence length components at
zero temperature. With the linearly extrapolated values of the upper
critical field at zero temperature, $\mu _{0}H_{\text{c2}}^{ab}\left(
0\right) =2.1(3)$ T and $\mu _{0}H_{\text{c2}}^{c}\left( 0\right) =0.25(2)$
T, and the equation above we obtain $\xi _{ab}\left( 0\right) =36.3(1.5)$ nm
and $\xi _{c}\left( 0\right) =4.3(7)$ nm. These values are comparable to
earlier reports such as resistivity measurements ($\xi _{ab}\left( 0\right)
=29.0$ nm and $\xi _{c}\left( 0\right) =5.7$ nm) \cite{critical_field}, or
susceptibility studies ($\xi _{ab}\left( 0\right) =35$ nm and $\xi
_{c}\left( 0\right) =13$ nm) \cite{Emery_susc}. Remarkably, the upper
critical field $H_{\text{c2}}^{ab}$ shows a positive curvature in the
studied temperature region which was not observed in previous
investigations. A similar positive curvature was observed near $T_{\text{c}}$
in MgB$_{2}$ \cite{mgb2poscurve} where it was explained by a two-gap model.
However, we should keep in mind that other explanations are also possible.
Complementary experiments, such as studies of the superfluid density, as
performed in this work, are necessary to draw definite conclusions. 
\begin{figure}[th]
\includegraphics[width=1\hsize]{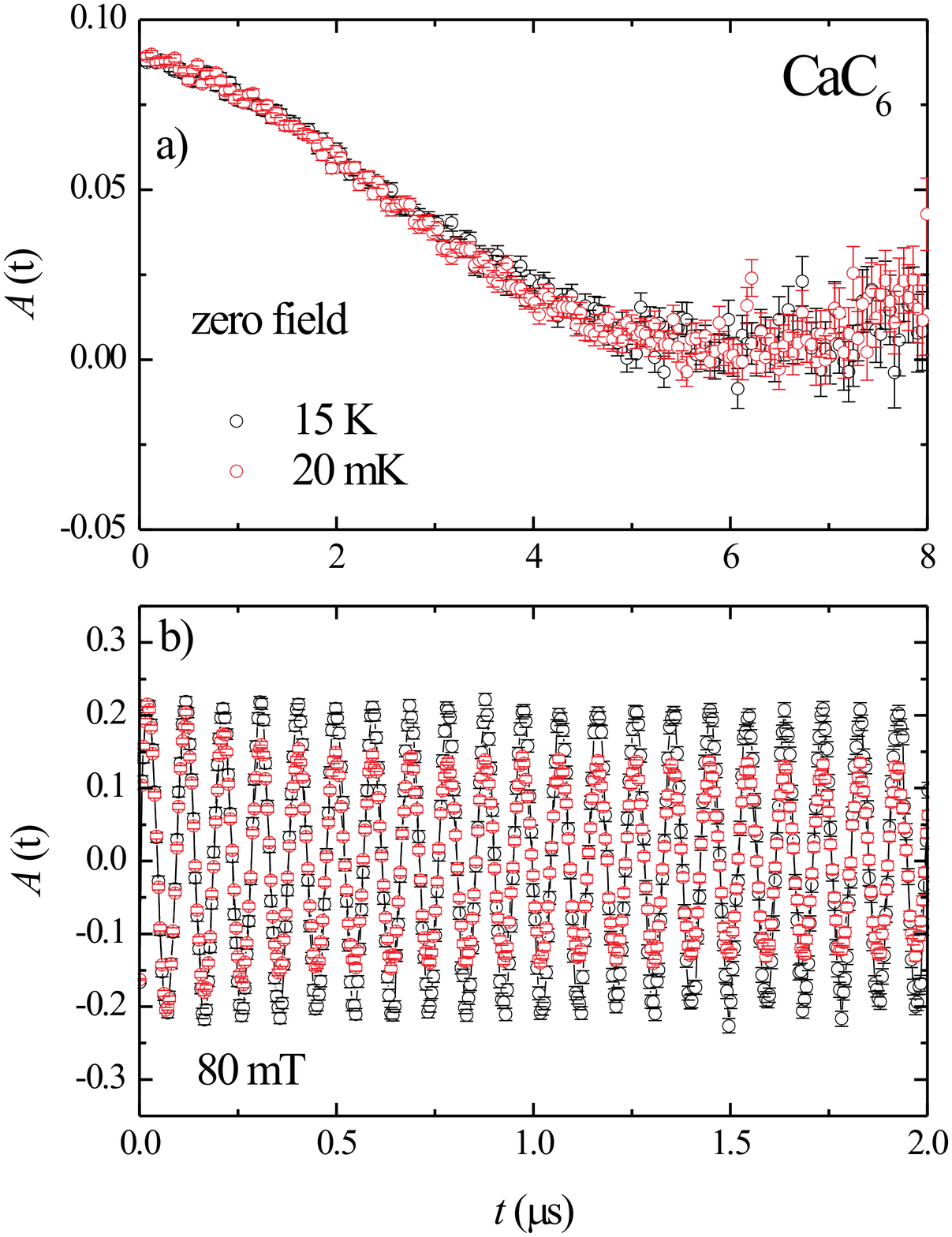}
\caption{(color online) Measured $\protect\mu $SR asymmetries $A(t)$ in CaC$%
_{6}$. a) Asymmetry in zero magnetic field (ZF). b) Asymmetry in 80 mT
transverse field (TF). The black data points were taken at 15 K (well above $%
T_{\text{c}}$) and the red ones at 20 mK.}
\end{figure}
The temperature dependence of the upper critical field anisotropy $\gamma
_{H}=$ $H_{\text{c2}}^{ab}/H_{\text{c2}}^{c}$ is shown in the inset of
Fig.~1b. We interpolated the measured upper critical field values with a
third order polynomial and determined the ratio $\gamma _{H}$. The
anisotropy paremeter $\gamma _{H}$ increases monotonically with decreasing
temperature, showing a similar temperature dependence as with MgB$_{2}$ \cite%
{mgb2torque}. Note that a single gap isotropic $s$-wave description of
superconductivity cannot account for this behaviour of $\gamma _{H}\left(
T\right) $.

$\mu $SR experiments were carried out at the $\pi $M3 beamline using the GPS
(General Purpose Spectrometer) and the LTF (Low Temperature Facility)
spectrometers at the Paul Scherrer Institute PSI Villigen, Switzerland. Six
pieces of CaC$_{6}$ forming an area of $\simeq $ 10x14 mm$^{2}$ were used in
the experiment. In the intercalation process the Ca atoms penetrate from the
side of the graphite sample and diffuse along the $ab$-plane. Reducing the
sample size in the $ab$-direction favors Ca diffusion, and consequently
leads to a higher sample quality than just using one large piece. For the
LTF experiments the samples were glued on a silver plate with the help of
Apiezon N grease and covered by silver coated polyester foil. In contrast,
for the GPS experiments the samples were supported between the two arms of a
specially designed fork by means of kapton tape. This technique allows only
the incoming muons that stop in the sample to be counted, and therefore the
background signal is reduced. 
\begin{figure}[th]
\includegraphics[width=1\hsize]{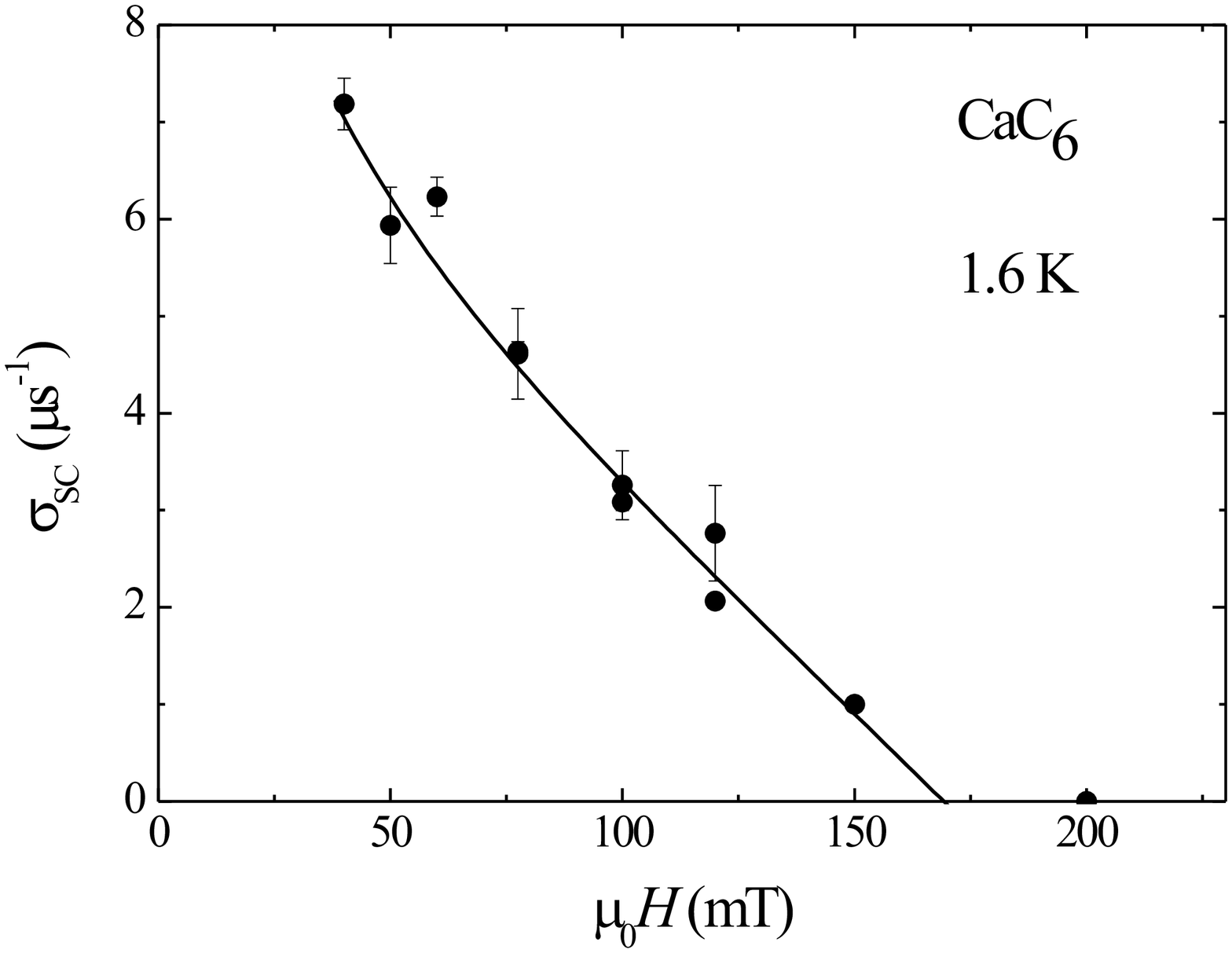}
\caption{Field dependence of the Gaussian relaxation rate $\protect\sigma _{%
\text{SC}}$ in CaC$_{6}$ measured at $T=1.6$ K. The solid line corresponds
to the calculated curve using Eq. (3) (see text).}
\end{figure}
The magnetic field (80 mT) was applied along the crystallographic $c$%
-direction at 15 K well above $T_{\text{c}}.$ Then the sample was cooled
down to the lowest temperature (20 mK for the LTF spectrometer), and $\mu $%
SR spectra were recorded with increasing temperature. The same procedure was
applied in the GPS spectrometer, down to the lowest temperature of 1.6 K
available in this device. The initial spin polarization of the implanted
muons is perpendicular to the applied magnetic field $H$. The $\mu $SR
experiments were performed at $H>H_{\text{c1}}^{\text{c}}$ so the sample was
in the vortex state. Magnetic field dependent measurements were performed in
the GPS spectrometer. At each magnetic field $H$ the sample was warmed up to
15 K (well above $T_{\text{c}}$), a $\mu $SR time spectrum was recorded,
then it was cooled down to 1.6 K, and another $\mu $SR time spectrum was
taken.\textbf{Before and after the }$\mu $\textbf{SR experiments the same
piece of CaC$_{6}$ was measured in the AC susceptometer, comparing the two
measurements we observed no difference.}

In the vortex state of a type II\ superconductor the muons probe the local
magnetic field distribution $P(B)$ due to the vortex lattice \cite%
{field_dist}. From the second moment $\langle \Delta B^{2}\rangle $ of $P(B)$
the in-plane magnetic penetration depth $\lambda _{ab}$\ can be extracted
according to the relation $\langle \Delta B^{2}\rangle =C\lambda _{ab}^{-2}$%
\ ($C$ is a field dependent quantity) from which the superfluid density $%
\rho _{\text{s}}\propto \lambda _{ab}^{-2}$ can be determined \cite%
{Brandt_num}.\ Thus, $\mu $SR provides a direct method to measure the
superfluid density in the mixed state of a type II superconductor. Figure 2a
shows the $\mu $SR asymmetries $A(t)$ recorded in LTF in zero magnetic field
above $T_{\text{c}}$ and at 20 mK. The two time spectra overlap, indicating
the absence of any kind of magnetic order in the sample and in the sample
holder. The time dependence of the asymmetry can be described by the static
Kubo-Toyabe formula, showing only the presence of randomly distributed
nuclear magnetic moments, but no realization of magnetism. Figure 2b shows
the $\mu $SR asymmetries taken at 15 K well above $T_{\text{c}}\simeq 11.5$
K and at 20 mK in a field of 80 mT. A clear damping of the $\mu $SR signal
at 20 mK due to the presence of the vortex lattice is visible. For a
detailed description of $\mu $SR studies of the vortex state in type II\
superconductors see, e.g. \cite{field_dist}. The $\mu $SR time spectra were
fitted to the following expression: 
\begin{eqnarray}
A^{\text{TF}}(t) &=&A_{\text{SC}}e^{(\sigma _{\text{SC}}t)^{2}/2}\cos
(\gamma _{\mu }B_{\text{SC}}+\phi )  \notag \\
&&+A_{\text{bg}}e^{(\sigma _{\text{bg}}t)^{2}/2}\cos (\gamma _{\mu }B_{\text{%
bg}}+\phi ) \\
\ &&+A_{\text{Ag}}\cos (\gamma _{\mu }B_{\text{Ag}}+\phi )  \notag
\end{eqnarray}%
\newline
Here the indices SC, bg and Ag denote the sample (superconductor), the
background arising from non superconducting parts of the sample and the
non-relaxing silver background, respectively. $A$ denotes the initial
asymmetry, $\sigma $ is the Gaussian relaxation rate, $\gamma _{\mu }=2\pi
\cdot 135.5342$~MHz/T is the muon gyromagnetic ratio, $B$ is the internal
magnetic field, and $\phi $ is the initial phase of the moun-spin ensemble.
A set of $\mu $SR data were fitted simultaneously with $A_{\text{SC}}$, $A_{%
\text{bg}}$, $A_{\text{Ag}}$, $B_{\text{bg}}$, $B_{\text{Ag}}$, $\sigma _{%
\text{bg}}$, and $\phi $ as common parameters, and $B_{\text{SC}}$ and $%
\sigma _{\text{SC}}$ as free parameters for each temperature. The same
expression was used to analyse the data recorded with the GPS spectrometer
(where no silver background signal is present). To analyze the magnetic
field dependent $\mu $SR measurements $B_{\text{SC}}$\ was a free parameter
in Eq. (2) since it depends on the applied field.

The analysis of the field dependence of the measured $\mu $SR relaxation
rate $\sigma _{\text{SC}}\propto $ $\langle \Delta B^{2}\rangle $ allows to
estimate the absolute value of the in-plane magnetic penetration depth $%
\lambda _{\text{ab}}$, assuming that CaC$_{6}$ can be described as a single
isotropic $s$-wave gap superconductor. The measured values of $\sigma _{%
\text{SC}}$\ are plotted in Fig.~3. To analyze our data we used the formula
developed by Brandt \cite{Brandt_num}: 
\begin{equation}
\sigma _{\text{SC}}\simeq 0.172\gamma _{\mu }\frac{\Phi _{0}}{2\pi }\left( 1-%
\frac{B}{B_{\text{c2}}^{c}}\right) \left[ 1+1.21\left( 1-\sqrt{\frac{B}{B_{%
\text{c2}}^{c}}}\right) ^{3}\right] \lambda _{ab}^{-2}\text{.}
\end{equation}

Although the formula was derived for high $\kappa $\ type II superconductors
($\kappa \geq 5$), it was suggested to work also for low values of $\kappa $%
\ at magnetic fields $\sqrt{B/B_{\text{c2}}^{c}}\geq 0.5$ \cite{DiCastro}.
From the measured $\mu $SR relaxation rates $\sigma _{\text{SC}}$ and Eq.
(3) one obtains $\lambda _{ab}\left( 0\right) =78(3)$ nm and $B_{\text{c2}%
}^{c}(0)=170(5)$ mT. The solid line in Fig.~3 represents the corresponding
fit (the point at 200 mT was not included in the fit). To estimate the
Ginzburg-Landau parameter $\kappa $ we use our values of $\xi _{ab}\left(
0\right) =36.3(1.5)$ nm and $\lambda _{ab}\left( 0\right) =78(3)$ nm giving $%
\kappa \approx 2.1$. It is worth to mention that the value of $B_{\text{c2}%
}^{\text{c}}$ is smaller than that estimated from the AC magnetization
measurements ($B_{\text{c2}}^{c}(0)\simeq 250$ mT), \textbf{since AC
magnetization measurements resulted more data points to determine the upper
critical field value we take 250 mT as $B_{\text{c2}}^{c}(0)$ for the
further analysis.} 
\begin{figure}[th]
\includegraphics[width=1\hsize]{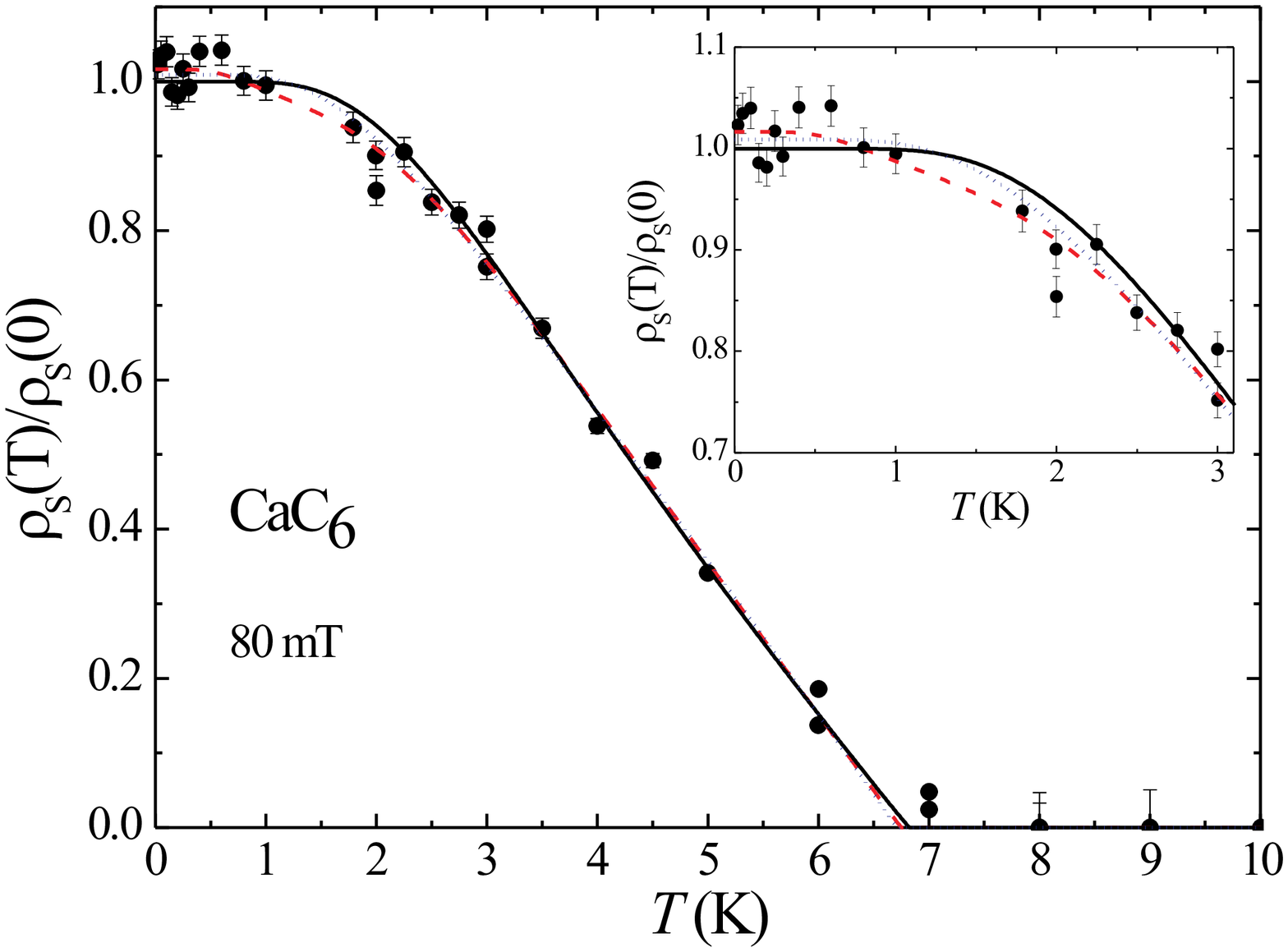}
\caption{(color online) Temperature dependence of the normalized superfluid
density $\protect\rho _{\text{s}}(T)/\protect\rho _{\text{s}}(0)$ in CaC$%
_{6} $ measured at 80 mT. The inset shows the low temperature region. The
black line corresponds to a single-gap isotropic $s$-wave, the blue, dotted
line to a single-gap anisotropic $s$-wave, and the red, dashed line to a
two-gap isotropic $s$-wave description of $\protect\rho _{\text{s}}(T)$. For
a detailed explanation see text.}
\end{figure}

It is generally assumed that the superfluid density $\rho _{\text{S}}$\ is
related to the in-plane magnetic penetration depth $\lambda _{\text{ab}}$\
by the simple relation $\rho _{\text{S}}\propto \lambda _{\text{ab}}^{-2}$.
However, for magnetic fields close to $B_{\text{c2}}$ this proportionality
must be corrected because the order parameter $\psi \left( r\right) $ is
reduced due to the overlapping vortices. In this case the spatial average of
the superfluid density is given by \cite{DiCastro}:

\begin{equation}
\rho _{\text{s}}\left( T\right) \simeq \langle |\psi \left( r\right)
|^{2}\rangle \lambda ^{-2}\left( T\right) \simeq \left( 1-B/B_{\text{c2}%
}^{c}(T)\right) \lambda _{ab}^{-2}\left( T\right) ,
\end{equation}%
where $\langle ...\rangle $ means the spatial average. Values of $\lambda
_{ab}^{-2}$ \ were calculated using Eq. (3) and for the values of $B_{\text{%
c2}}^{c}$ we used the results of our AC magnetization measurements (see
Fig.~1b). The temperature dependence of the superfluid density $\rho _{\text{%
s}}\left( T\right) $ determined at 80 mT is plotted in Fig.~4. We analyzed $%
\rho _{\text{s}}\left( T\right) $ using three different models: \textit{i)}
single-gap isotropic $s$-wave \textit{ii)} single-gap anisotropic $s$-wave,
and \textit{iii)} two-gap isotropic $s$-wave. To calculate the temperature
dependence of the magnetic penetration depth within the local approximation (%
$\lambda \gg \xi $) we applied the following formula \cite{gap1,gap2}: 
\begin{equation}
\rho _{\text{s}}\left( T\right) =\rho _{\text{s}}\left( 0\right) \left( 1+%
\frac{1}{\pi }\int_{0}^{2\pi }\int_{\Delta \left( T,\phi \right) }^{\infty
}\left( \frac{\partial f}{\partial E}\right) \frac{E\text{ }dEd\phi }{\sqrt{%
E^{2}-\Delta \left( T,\phi \right) ^{2}}}\right) \text{,}
\end{equation}%
where $\rho _{\text{S}}\left( 0\right) $ is the zero temperature value of
the superfluid density, $f=\left[ 1+\exp \left( (E-E_{\text{F}})/k_{\text{B}%
}T\right) \right] ^{-1}$ is the Fermi function, $\phi $ is the angle along
the Fermi surface, and $\Delta \left( T,\phi \right) =\Delta _{0}h\left(
T/T_{\text{c}}\right) g\left( \phi \right) $. The temperature dependence of
the gap is expressed by $h\left( T/T_{\text{c}}\right) =\tanh \left( 1.82%
\left[ 1.018\left( T_{\text{c}}/T-1\right) \right] ^{0.51}\right) $ \cite%
{gap3}. For the isotropic $s$-wave gap \textit{i)} $g^{\text{s}}\left( \phi
\right) =1$ and for the anisotropic $s$-wave gap \textit{ii)} $g^{\text{s}_{%
\text{An}}}\left( \phi \right) =\left( 1+a\cos 4\phi \right) /\left(
1+a\right) $, where $a$ denotes the anisotropy of the gap. The two-gap fit
was calculated using the so called $\alpha $-model and assuming that the
total superfluid density is the sum of the two components \cite{gap2,gap3}: 
\begin{equation}
\rho _{\text{s}}\left( T\right) =\rho _{\text{s}}\left( 0\right) \left(
\omega \frac{\rho \left( T,\Delta _{0,1}\right) }{\rho \left( 0,\Delta
_{0,1}\right) }+\left( 1-\omega \right) \frac{\rho \left( T,\Delta
_{0,2}\right) }{\rho \left( 0,\Delta _{0,2}\right) }\right) \text{.}
\end{equation}

Here $\Delta _{0,1}$ and $\Delta _{0,2}$ are the zero temperature values of
the larger and the smaller gap, and $\omega $ $\left( 0\leq \omega \leq
1\right) $ is a weighting factor representing the relative contribution of
the larger gap to $\rho _{\text{s}}$. The results obtained for the different
models are plotted in Fig.~4. The black line represents the result for the
single $s$-wave gap model \textit{i)} with $\rho _{\text{s}}\left( 0\right)
=96.5(5)$ $\mu $m$^{-2}$, $\Delta _{0}=0.79(1)$ meV, $T_{\text{c}}=6.80(2)$
K, and $2\Delta _{0}/k_{\text{B}}T_{\text{c}}=2.70(5)$, describing CaC$_{6}$
as weak coupling BCS superconductor. The reduction of $T_{\text{c}}$ agrees
well with our AC magnetization measurements for 80 mT. The blue, dotted line
represents the anisotropic $s$-wave gap analysis \textit{ii) }with $\Delta
_{0}=0.81(1)$ meV and $a\simeq 0.37$ for the gap anisotropy. These values do
not agree with the reported ones from tunneling experiments ($\Delta
_{0}^{c}\simeq 1.7$ meV and $\Delta _{0}^{ab}\simeq 1.44$ meV) \cite%
{pointcontact}. The two-gap analysis \textit{iii)} (red, dashed line) yields 
$\Delta _{0,1}=0.85(2)$ meV for the larger gap and $\Delta _{0,2}=0.23(1)$
meV for the smaller one. Only $\sim 9.4(3)$ \% of the superfluid density is
associated with the smaller gap. The smaller gap agrees with $\Delta
_{0,2}=0.2(2)$ meV reported in \cite{arpes}. Our larger gap value is
significantly smaller ($\Delta _{0,1}=1.9(2)$ meV). \textbf{It is worth to
mention that the applicability of the two gap model at this magnetic field
(80 mT) is questionable. The gap magnitude scales with the upper critical
field, $\left( \frac{\Delta _{0,1}}{\Delta _{0,2}}\right) ^{2}\sim \frac{B_{%
\text{c2}}^{1}}{B_{\text{c2}}^{2}},$ resulting $B_{\text{c2}}^{2}\sim 12$ mT
which is much smaller than the applied magnetic field.} Although the best
agreement is found for the two-gap scenario, no final conclusion can be
drawn from this analysis. All three scenarios describe well the observed
temperature dependence of the superfluid density. For a direct comparison
with Ref. \cite{DiCastro} we omitted the values of $\rho _{\text{s}}\left(
T\right) $ below 2 K. In this case the single s-wave gap fit yields $\rho _{%
\text{s}}\left( 0\right) =91.6(9)$ $\mu $m$^{-2}$, $\Delta _{0}=0.84(1)$
meV, $T_{\text{c}}=6.75(2)$ K, and $2\Delta _{0}/k_{\text{B}}T_{\text{c}%
}=2.90(5)$. The values presented in Ref. \cite{DiCastro} are: $\rho _{\text{s%
}}\left( 0\right) =85.1(3)$ $\mu $m$^{-2}$, $\Delta _{0}=0.868(5)$ meV, and $%
2\Delta _{0}/k_{\text{B}}T_{\text{c}}=3.6(1)$. The higher applied magnetic
field (120 mT in Ref. \cite{DiCastro}) may account for the differences.
Here, one should point out that our low temperature data are essential to
determine the zero temperature values of $\rho _{\text{s}}$\ and $\Delta $.

\section{Conclusions}

In conlusion, we carried out extensive AC magnetization measurements to map
the temperature dependence of the upper critical field $H_{\text{c2}}^{ab}$
and $H_{\text{c2}}^{c}$ in CaC$_{6}$ and to evaluate the coherence length.
The values of $\xi _{ab}\left( 0\right) =36.3(1.5)$ nm and $\xi _{c}\left(
0\right) =4.3(7)$ nm are in good agreement with those reported previously 
\cite{critical_field,Emery_susc}. We found that the upper critical field
anisotropy $\gamma _{H}=$ $H_{\text{c2}}^{ab}/H_{\text{c2}}^{c}$ increases
with decreasing temperature. From magnetic field dependent $\mu $SR
experiments the absolute value of the in-plane magnetic penetration depth
was determined to be $\lambda _{ab}\simeq 78(3)$ nm, in agreement with
previously reported values \cite{LamuraPRL,DiCastro}. Furthermore, low
temperature $\mu $SR experiments were performed in order to map the whole
temperature dependence of the superfluid density $\rho _{\text{s}}\left(
T\right) $. We analyzed the temperature dependence of $\rho _{\text{s}}$
with three different models: \textit{i)} single-gap isotropic $s$-wave, 
\textit{ii)} single-gap anisotropic $s$-wave, and \textit{iii)} two-gap
isotropic $s$-wave. All models describe the measured $\mu $SR data almost
equally well, although a slightly better agreement was achieved using the
two-gap model.

This work was partly supported by the Swiss National Science Foundation
(SCOPES grant No. IZ73Z0\_128242). The $\mu $SR experiments were performed
at the Swiss Muon Source, Paul Scherrer Institut, Villigen, Switzerland.


\end{document}